\documentclass[twocolumn, 11pt]{article}

\usepackage[backend=biber,style=ieee]{biblatex}
\usepackage{graphicx}
\usepackage{dcolumn}
\usepackage{hyperref}
\usepackage{multirow}
\usepackage{textalpha}
\usepackage{color,soul} %
\usepackage{verbatim}
\usepackage{booktabs}

\usepackage{xr}
\usepackage{xr-hyper}
\usepackage{hyperref} %

\definecolor{hotpink}{RGB}{255, 92, 167}

\newcommand{\newtextcolor}{black}  %
\newcommand{\newertextcolor}{black}  %

\begin{document}

\twocolumn
\title{MDIntrinsicDimension: Dimensionality-Based Analysis of Collective Motions in Macromolecules from Molecular Dynamics Trajectories}

\author{Irene Cazzaniga\textsuperscript{1}, Toni Giorgino\textsuperscript{1,*}}
\date{\small \textsuperscript{1}Istituto di Biofisica (IBF-CNR), Consiglio Nazionale delle Ricerche, 20133 Milano, Italy.\\ \ *\  E-mail: \texttt{toni.giorgino@cnr.it}}

\maketitle

\begin{abstract}
Molecular dynamics (MD) simulations provide atomistic insights into the structure, dynamics, and function of biomolecules by generating time-resolved, high-dimensional trajectories. Analyzing such data benefits from estimating the minimal number of variables required to describe the explored conformational manifold, known as the intrinsic dimension (ID). We present \texttt{MDIntrinsicDimension}, an open-source Python package that estimates ID directly from MD trajectories by combining rotation- and translation-invariant molecular projections with state-of-the-art estimators. The package provides three complementary analysis modes: whole-molecule ID; sliding windows along the sequence; and per–secondary-structure elements. It computes both overall ID (a single summary value) and instantaneous, time-resolved ID that can reveal transitions and heterogeneity over time. We illustrate the approach on fast folding-unfolding trajectories from the DESRES dataset, demonstrating that ID complements conventional geometric descriptors by highlighting spatially localized flexibility, differentiating structural segments, and identifying a metastable configuration. 
\end{abstract}

\section{Introduction}
Molecular dynamics (MD) simulation is a computational technique that provides time-resolved, atomistic descriptions of biomolecules by integrating the corresponding equations of motion \cite{karplus_molecular_1990,hollingsworth_molecular_2018}. The resulting trajectories are high-dimensional and often challenging to interpret directly. While dimensionality reduction techniques embed such data into lower-dimensional spaces \cite{tribello_using_2012,tsai_sgoop_2021}, an important question is how many variables are minimally required to  describe the underlying data manifold, that is, its intrinsic dimension~(ID) \cite{camastra_intrinsic_2016,campadelli_review_2015}.

\textcolor{\newtextcolor}{From a geometric perspective, the ID of a trajectory can be interpreted as the effective number of independent collective coordinates that are needed to describe the \textit{important} fluctuations actually sampled by the system. In contrast to the formal number of atomic degrees of freedom, which includes many frozen or highly correlated directions, the ID discounts those redundant motions and therefore directly reflects how many modes of motion are relevant on the timescale and in the coordinate representation under study. \cite{del_giudice_effective_2021}}

Estimating the ID of all-atom trajectories is a challenging and arguably ill-defined task due to several factors. First, the conformational space of biomolecules is inherently high-dimensional, which amplifies data sparsity. Second, meaningful internal degrees of freedom must be distinguished from irrelevant components such as noise and rigid-body motions. Third, the sampling density is not uniform, both across different regions of conformational space and over time \cite{facco_estimating_2017}. The latter is a particularly important consideration for proteins, whose flexibility varies locally and on multiple timescales \cite{das_low-dimensional_2006}.

Here, we introduce \texttt{MDIntrinsicDimension}, an open-source Python package that estimates ID directly from MD trajectories. The package computes internal, rotation- and translation-invariant projections (e.g., backbone dihedrals and inter-residue distances) and applies modern ID estimators provided by the \texttt{scikit-dimension} package \cite{bac_scikit-dimension_2021}. Three analysis modes are provided, namely  whole-molecule, sliding windows along the sequence; and secondary structure elements (Table~\ref{tab:modes}). These complementary views enable both system-level summaries and spatially localized assessments of flexibility.

Orthogonally, the package provides three time-resolved representations of ID,  namely: \textit{overall}, \textit{instantaneous} and \textit{averaged}. Overall ID summarizes the dataset with a single value along the trajectory; instantaneous ID is time-resolved, providing a per-frame estimate that can be averaged over the full trajectory or over trailing segments to detect transitions  (Table~\ref{tab:modes}). Together, these outputs characterize the complexity of the explored conformational space and its temporal evolution.

We demonstrate our approach on trajectories from the D.\ E.\ Shaw Research (DESRES) fast-folding protein trajectory set \cite{lindorff-larsen_how_2011}, including  the Nle/Nle double mutant of the HP35 C-terminal fragment of the villin headpiece (henceforth ``villin'') and N-terminal Domain of Ribosomal Protein L9 (NTL9), and we show that ID complements conventional geometric measures by highlighting heterogeneity across residues and secondary structure elements.

\section{Methods}

The ID calculation workflow comprises three stages: (i) load a MD trajectory; (ii) compute an internal, rigid-body–invariant, projection chosen between the available options, to obtain a frame-by-feature matrix; (iii) estimate the ID proper using an  estimator chosen out of a set of available methods.

\begin{table*}
    \centering \small
    \begin{tabular}{lllc}
    \toprule
    \textbf{Locality} &\bf Definition & \bf{Meaning} &  \bf Call \\
    \midrule
       \multirow{3}{3em}{\bf{Sequence}} & Whole protein & Protein-wide value & \verb+intrinsic_dimension()+ \\
       & Section & ID by overlapping sliding windows & \verb+section_id()+ \\
       & Secondary structure & ID of secondary structure elements & \verb+secondary_structure_id()+ \\
    \midrule
     \multirow{3}{4.5em}{\bf{Time}} &Instantaneous & Time-resolved via local estimators* & \\ 
       & Averaged & Mean of instantaneous ID along a trajectory & \\
       & Overall & ID of the trajectory  via global estimators* \\
    \bottomrule
    \end{tabular}
    \caption{Nomenclature on temporal and spatial ID types adopted in this package. *Local estimators can be implicitly converted into global ones and vice-versa.
    }
    \label{tab:modes}
\end{table*}

\subsection{Internal coordinate projections}

Estimating structural properties from molecular dynamics (MD) trajectories benefits from representations that are invariant to rigid-body motions, thereby reflecting only the molecule’s internal degrees of freedom. To this end, each frame of a trajectory is mapped to a vector of $m$ descriptors using suitable projection functions. The resulting projections are stored as an  $n \times m$ matrix, where each of the  $n$  rows corresponds to a trajectory frame and each of the $m$ columns to a structural feature.

In practice, we employ two complementary families of descriptors: (i) inter-residue distances or contact counts, typically computed between Cα atoms, which emphasize medium- and long-range couplings within the structure; and (ii) torsional angles, including backbone (\(\phi, \psi\)) and, when relevant,  side-chain $(\chi)$  dihedrals, which capture local conformational variability along the chain.

Trajectory handling and projection calculations are performed using the MoleculeKit library \cite{doerr_htmd_2016}. Unless otherwise specified, Cα atoms are used for distance calculations, and dihedral angles are expressed in degrees. Periodicity of angular variables can be handled through a sine–cosine embedding. Arbitrary projection schemes can also be employed, provided they map each frame to a real-valued feature vector.

\subsection{Intrinsic dimension estimation} \label{ID types}

\textcolor{\newtextcolor}{To provide an intuition, different classes of methods estimate the dimensionality of the underlying manifold by expressing the same core idea through complementary formalizations. Broadly speaking, distance-based methods look at how the number of neighbours grows as one increases the distance around each point (frame), that is, the power law of how the ``volume’’  of configuration space increases with radius \cite{glielmo_dadapy_2022}. Other approaches examine principal axes and how their associated variances decay, or related constructions. In all cases, the ID summarizes how many effectively independent directions the system explores   and thus plays a role loosely analogous to the number of principal components or time-lagged independent components needed to   capture most of the variance in a linear embedding, while being defined in a non-linear way and allowing for variations in the  instantaneous ID. \cite{campadelli_review_2015}}

This package estimates ID using any of the algorithms implemented in \texttt{scikit-dimension} \cite{bac_scikit-dimension_2021}, which notably include nearest-neighbour, fractal and likelihood-based operating principles (Supplementary Table~\ref{tab:scikit_estimators}).  We found the Two Nearest Neighbours (\textit{TwoNN}) estimator  \cite{facco_estimating_2017} to be robust, fast and well-behaved on MD data, and we adopt it as the default, unless otherwise specified.

Three complementary summary metrics are reported. First, an \textit{overall} estimate which treats the full set of frames as a single point cloud and yields a single ID value; to mitigate the influence of possibly non-equilibrated  initial segments we additionally report a overall estimate computed on the final portion of the trajectory (whose  length can be specified as a parameter). Second, an \textit{instantaneous} estimate produces a time series by evaluating ID in neighbourhoods centred at each frame. Third, we summarize this instantaneous series by its mean over the full trajectory and over its final portion (\textit{averaged}).  In summary, the \textit{instantaneous} ID consists in a vector with as many elements as there are frames in the trajectory, while \textit{averaged} and \textit{overall} ID are real numbers characterizing the whole trajectory. \textcolor{\newertextcolor}{\textit{Averaged} and \textit{overall} ID, while conceptually distinct, were found to yield similar values in our tests}  (Supplementary Figure \ref{fig:global id}).

\subsection{Localizing ID along sequence and structure}
The package enables space-localized structural analyses to uncover locality and heterogeneity in conformational complexity. First,  a sliding-window scheme partitions the protein sequence in overlapping windows of fixed length and stride;  for each window we recompute the internal coordinate projection restricted to the atoms in that window, followed by ID estimation. This produces a profile of ID along the primary sequence, highlighting different types of large-scale coordinated flexibility.

\textcolor{\newtextcolor}{Because neighbouring windows share most of their residues, the corresponding ID estimates are strongly correlated. They are best interpreted as a smooth mesoscale profile along the sequence, with an effective spatial resolution set by the window length (and, to a lesser extent, the stride), rather than as independent per-residue measurements. In short, the window should be large enough that the dimensionality of the original embedding space is much larger than the ID, yet small enough to retain local information; this trade-off will be demonstrated in detail with reference to villin trajectories in section \ref{sec:villin-id-by-sequence} and Supplementary Figures \ref{fig:villin windows} and \ref{fig:villin strides} (refer to the corresponding captions). On the other hand, increasing the stride has no effect on the overall trends but  can  speed-up the analysis by omitting some windows.}

Alternatively, for a structure-based scheme, the package can assign secondary structure using DSSP \cite{kabschDictionaryProteinSecondary1983, hekkelman_dssp_2025}, grouping consecutive residues sharing the same assignment into  segments.  For each segment,  projections and ID values are computed as above. We employ the simplified  coil (C), strand (E) and helix (H) DSSP scheme unless stated otherwise; results with the full DSSP alphabet are consistent.

The analysis returns, for each window or secondary structure element, either the overall summary as defined above, or the instantaneous series and the corresponding averaged summary, enabling direct comparisons across sequence positions and structural types.

\section{Results} \label{sec:results}
\subsection{Case Study: Villin Headpiece}
\begin{figure}
\centering
\includegraphics[width=0.9\linewidth]{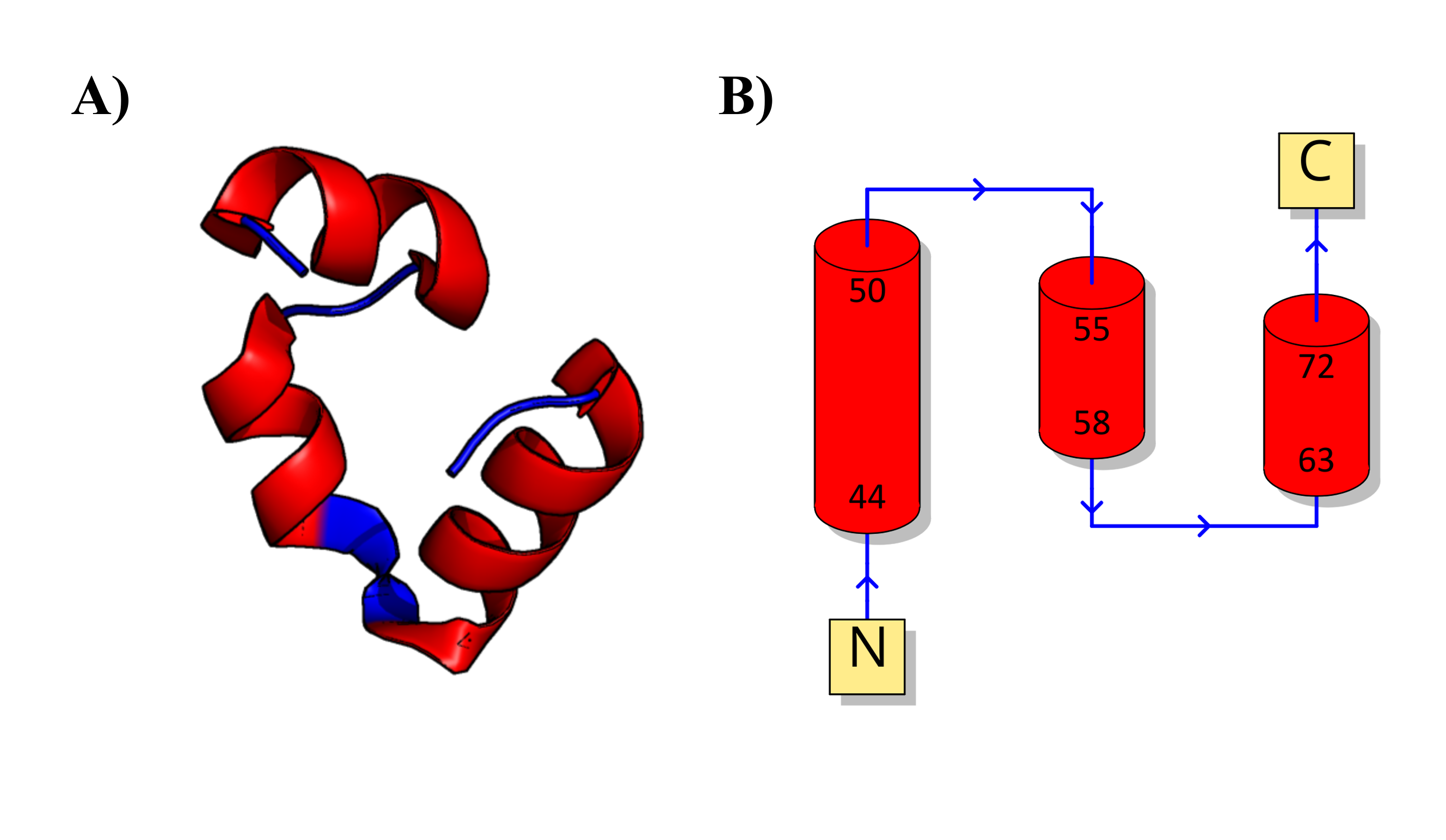}
    \caption{(A) Structure of the HP35 villin headpiece (PDB 2F4K) used in the case study and (B)  topology diagram.}
    \label{fig:villin_topology}
\end{figure}

We demonstrate the features of our package using trajectories of spontaneous folding-unfolding transitions from the dataset of fast-folding proteins provided by  DESRES in \cite{lindorff-larsen_how_2011}. We selected the HP35 chicken villin headpiece (PDB: 2F4K), a 35 residue  protein with two point mutations to norleucine, K65(NLE) and K70(NLE), which increase the folding rate up to five-fold compared to the wild type (Figure~\ref{fig:villin_topology}).
The original trajectory was evaluated using RMSD relative to the folded structure (reference frame 10,400). Based on this analysis, we selected six trajectory segments of 200 ns each (2,000 frames), representing the protein either in folded or unfolded states. These segments are labelled as \verb|f0|, \verb|f1|, \verb|f2| for the folded state and \verb|u0|, \verb|u1|, \verb|u2| for the unfolded state, as per Supplementary Table~\ref{tab: villin trajectory}.

\subsubsection{Estimator performance}
We first evaluated all the estimators provided by \texttt{scikit-dimension} \cite{bac_scikit-dimension_2021} to asses their ability to handle the complexity of data derived from MD trajectories. Three main aspects were taken in account: (i) the capability to clearly differentiate between the two states (i.e. folded and unfolded); (ii) the ability to distinguish real degrees of freedom from noise; (iii) computation efficiency. 

\begin{figure}[t]
    \centering
    \includegraphics[width=0.9\linewidth]{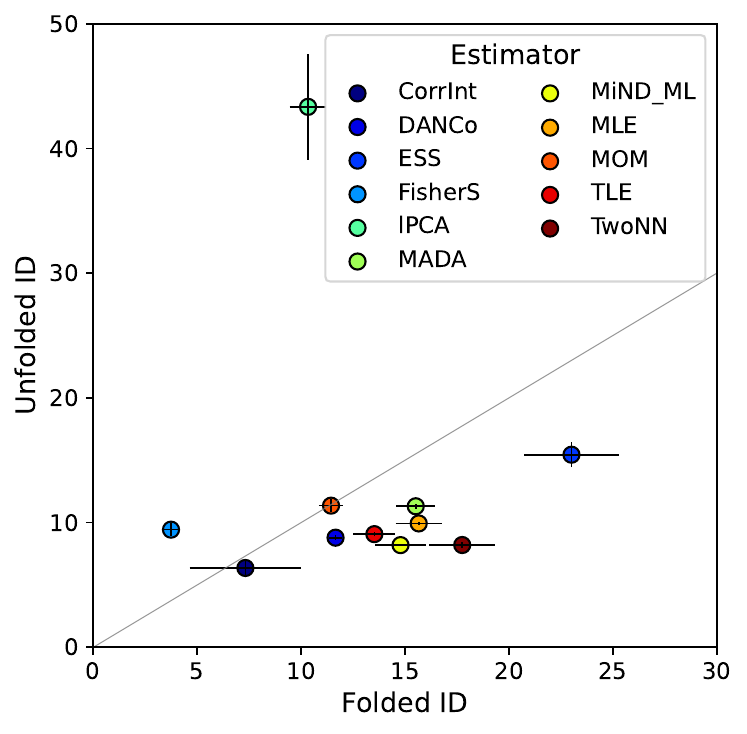}
    \caption{Folded-vs-unfolded ID of the villin dynamic manifold, computed by the estimators available at \texttt{scikit-dimension} package. Each point represents the mean value of the folded states ($x$ axis) and unfolded states ($y$ axis); error bars indicate standard deviations over three replicas.  Methods abbreviations are in Supplementary Table~\ref{tab:scikit_estimators}.  Projection: Ramachandran angles $\phi$ and $\psi$, whole protein. \textit{KNN} not  shown due to its high variance.}
    \label{fig:estimators comparison}
\end{figure}

All estimators were able to compute ID on the protein models used in this article (Figure \ref{fig:estimators comparison}).
Most of the remaining estimators succeed in discriminating between folded and unfolded states, generally reporting higher ID values for the folded case. The absolute ID values vary considerably across estimators, showing different discrimination effectiveness between the two states; for example \textit{MOM} reported very similar values for the two states;  \textit{lPCA} consistently reports the highest ID values measured from the unfolded trajectories (ID $> 35$).

Based on comparisons across different projections, conditions and  models, we selected \textit{TwoNN} as the default estimator in our functions, as it offers the best compromise between accuracy and computational cost.

\subsubsection{Projections}

For the sake of  illustration,  we focus on two main classes of projections, namely carbon–carbon distances (computed either on C$\alpha$ or C$\beta$ atoms), and torsional angles. For the latter, we distinguished between backbone dihedrals ($\phi$, $\psi$), which primarily describe backbone flexibility, and side-chain dihedrals ($\chi$), which probe side-chain conformational variability. Both types of torsion were considered in their raw angular form as well as in their sine/cosine embeddings to account for  periodicity.

\begin{figure}
    \centering
    \includegraphics[width=\linewidth]{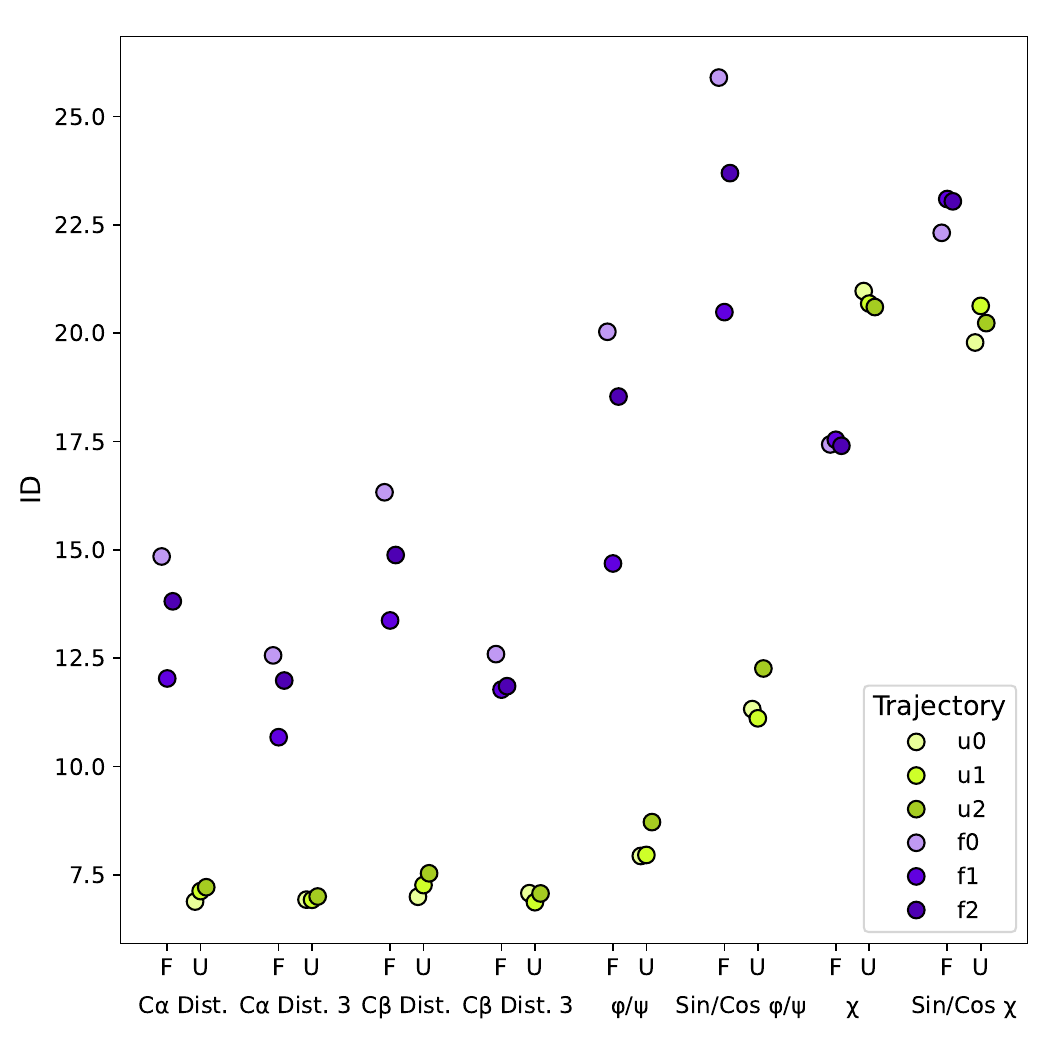}
    \caption{ID shifts between folded (F) and unfolded (U) states of villin under different projections.  \textit{Dist.}:  pairwise distances between all carbon--carbon pairs; \textit{Dist.\ 3}: pairwise distances every 3rd carbon; $\phi, \psi$: Ramachandran angles; $\chi$: sidechain dihedrals; \textit{Sin/Cos}: trigonometric embedding of dihedrals. }
    \label{fig:projections}
\end{figure}

All the selected projections are capable of distinguishing between folded and unfolded states  (Figure~\ref{fig:projections}). However, the ID values obtained, as well as the gap between the two states, can differ substantially depending on the projection. Remarkably, however, in the case of $\chi$ dihedral angles, where the trend is inverted: the unfolded state is associated with a higher ID, in contrast to the other projection types, where the folded state typically shows the higher ID. 
We attribute this inversion to the fact that the number of $\chi$ angles is residue-dependent, which introduces variability in the side-chain conformational space that is more prominent in the unfolded ensemble.  

\textcolor{\newtextcolor}{In this sense, higher ID values obtained with $\chi$ angles can be interpreted as reporting increased heterogeneity in side-chain conformational space, whereas distance- and backbone-based projections mainly track changes in tertiary contacts and backbone shape. For practical applications, we therefore recommend distance or backbone-dihedral projections when the goal is to characterize global folding/unfolding behaviour or large-scale structural transitions, and $\chi$-based projections when one is specifically interested in side-chain plasticity, for example in binding interfaces or active sites.}

\subsubsection{Instantaneous, Averaged and Overall ID}
\begin{figure}
    \centering
    \includegraphics[width=.9\linewidth]{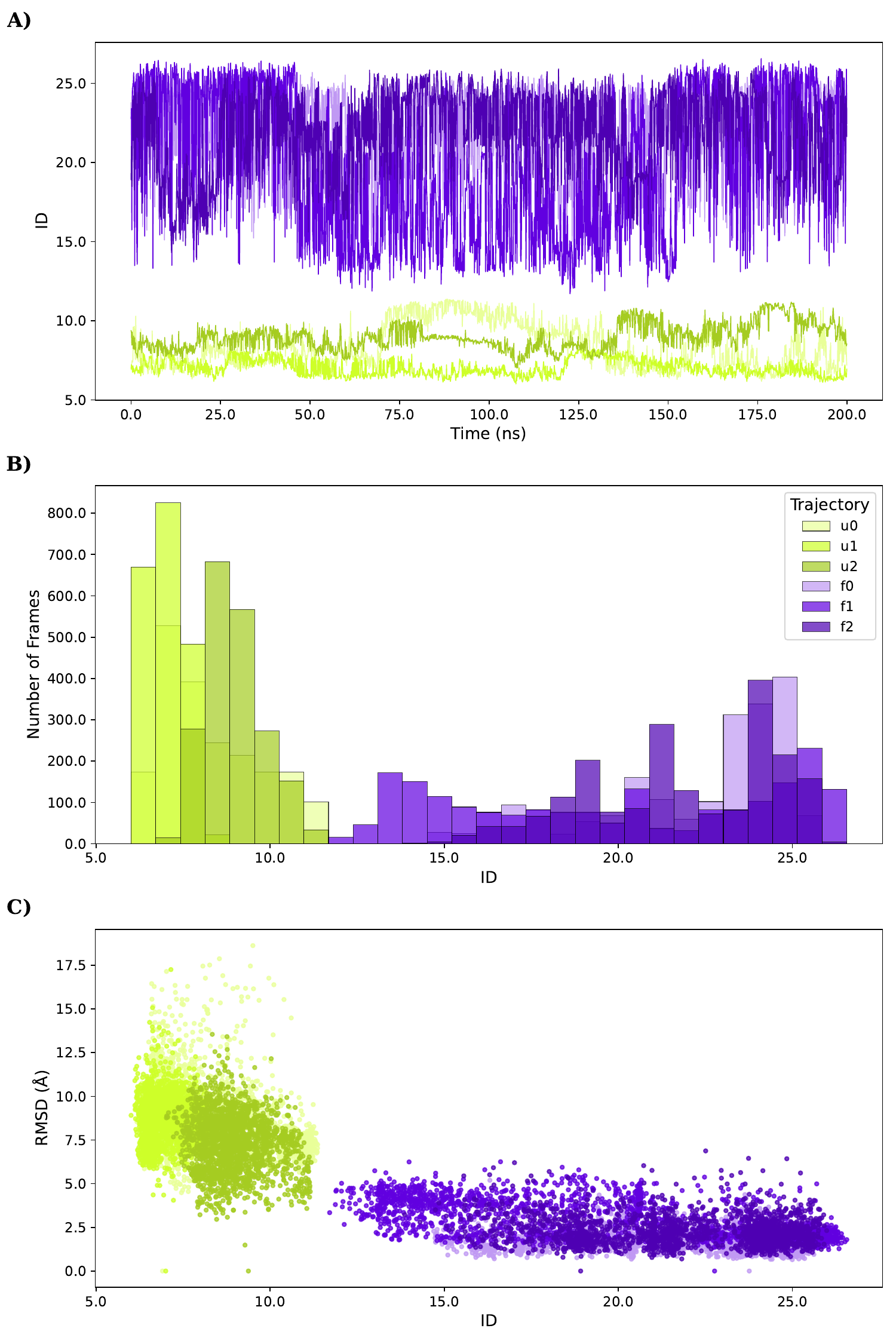} 
    
   \caption{(A) Instantaneous intrinsic dimension (ID) of villin trajectories over time. (B) Distribution of ID values across the trajectory. (C) Relationship between instantaneous ID and RMSD relative to the folded state. The separation between folded and unfolded ensembles is clearer when using ID than RMSD. States are color-coded as folded (violet) and unfolded (green). Analysis performed on the projection to $\phi$ and $\psi$ Ramachandran angles.}
    \label{fig:villin_instantaneous_and_RMSD}
\end{figure}

As described in Section~\ref{ID types}, \textit{instantaneous, averaged} and \textit{overall} ID are complementary approaches that provide distinct perspectives on system behavior: instantaneous ID captures frame-by-frame variations in dimensionality (Figure~\ref{fig:villin_instantaneous_and_RMSD}(A)), whereas the time-averaged and overall ID yield compact summaries that facilitate comparison across trajectories. Although these two summary measures may appear similar, they emphasize different aspects (Supplementary Figure~\ref{fig:global id}): the time-averaged ID, obtained as the mean of the instantaneous estimates, reflects the most frequently sampled conformations during the simulation. In contrast, the overall ID is computed directly from the pooled trajectory, treating it as a single dataset, and therefore quantifies the dimensionality of the conformational ensemble as a whole.

\subsubsection{Comparison with RMSD}

Root-mean-square deviation (RMSD) is commonly used to quantify structural changes in proteins when a reference structure is available. Accordingly, it is natural to compare instantaneous RMSD values with instantaneous ID values. When RMSD is computed relative to the folded reference structure, unfolded trajectories produce higher RMSD values, whereas ID assigns higher values to the folded trajectories.

This difference reflects the nature of the two metrics: RMSD measures how much a structure deviates from a fixed reference over time, whereas ID quantifies the effective number of degrees of freedom in the motion, determined by intramolecular constraints and the types of motions available. A compact folded globule -- without large flexible “unfolded hinges” -- appears to explore more effective degrees of freedom. Both metrics distinguish folded from unfolded trajectories (Figure~\ref{fig:villin_instantaneous_and_RMSD}(C)), but ID provides a sharper separation, with no overlap between the distributions.

\textcolor{\newtextcolor}{While RMSD was chosen as a widely used baseline observable,  ID can also be compared with alternative order parameters, such as ID estimated from PCA's fraction of explained variance (lPCA), or  projections of the trajectory on time-lagged independent components (tICA)  \cite{tica}. For both villin and NTL9, estimates based on the leading principal components (Supplementary Figures \ref{fig:villin lpca rmsd} and \ref{fig:ntl9 lpca rmsd}) or the first two time-lagged independent components (Supplementary Figures \ref{fig:villin tica rmsd} and \ref{fig:ntl9 tica rmsd}) yielded broader and partially overlapping distributions
between folded and unfolded states. In contrast, the instantaneous ID estimated by TwoNN produced much sharper separation with minimal overlap. These observations suggest
that, at least for the systems tested here, ID captures aspects of conformational heterogeneity that are not easily summarized
by linearly projected coordinates alone.}

\subsubsection{ID by Sequence} \label{sec:villin-id-by-sequence}

To gain higher-resolution insight into protein dynamics, we computed sequence-local and structure-local intrinsic dimensionality (ID).
In Figure~\ref{fig:section and ss}(A) we show the results of \texttt{section\_id()} applied on villin headpiece, partitioned into seven overlapping windows of 15 residues each, with a sliding step of three residues. Although the fixed window length and substantial overlap introduce a strong correlation between adjacent segments from the same trajectory, the ID profiles still reveal a clear distinction between folded and unfolded states, as well as minor variations among replicas within each window.

\textcolor{\newtextcolor}{We explored a range of window lengths and strides (5--30 residues and 1--4-residue strides) and found that the main trends distinguishing folded from unfolded segments and the location of high-ID regions were robust, while larger windows primarily smoothed out fine-scale variations in the profiles (Supplementary Figures \ref{fig:villin windows} and \ref{fig:villin strides}).}

\subsubsection{ID by Secondary Structure}

\begin{figure*}[t]
    \centering
    \includegraphics[width=0.9\linewidth]{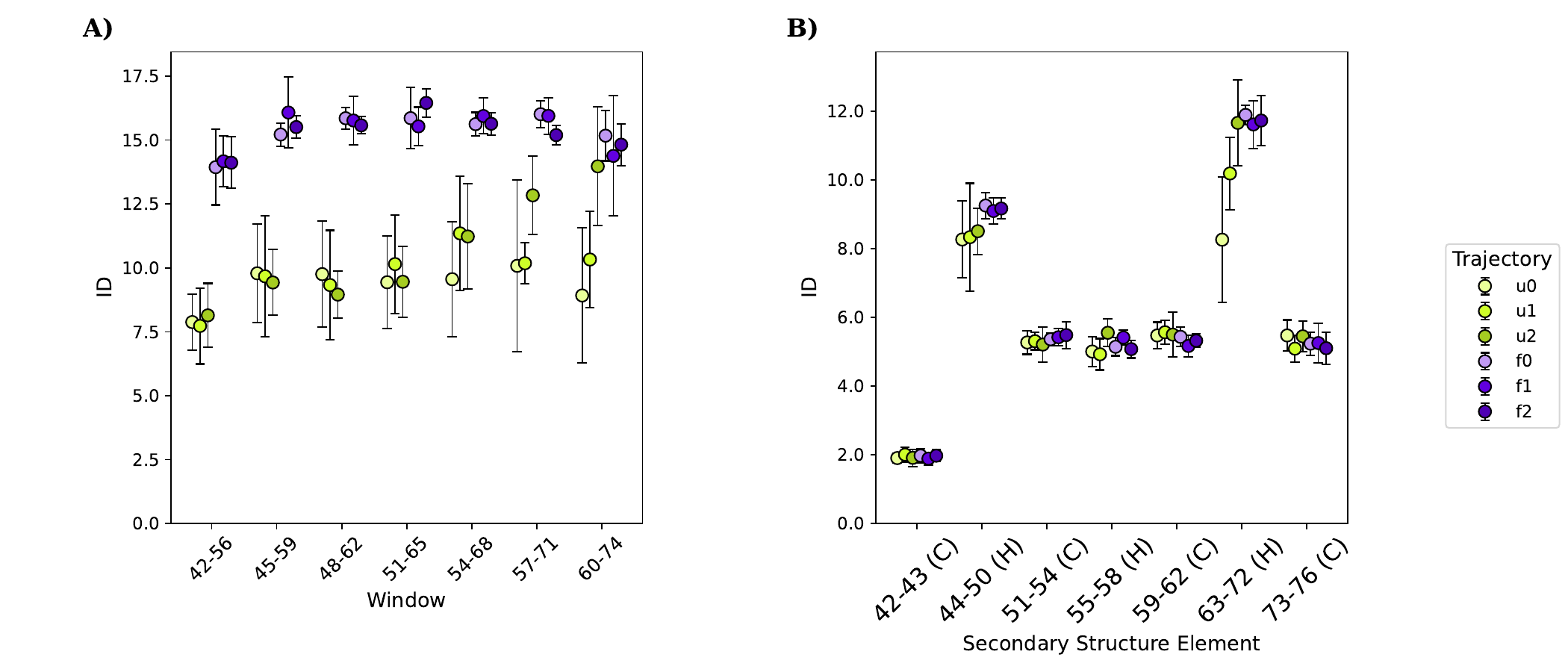}
    \caption{ID computed on local segments of villin. 
    (A) Sequence-wise ID computed with \texttt{section\_id()} using a window of 15 and stride of 3.
    (B) Secondary structure element-wise ID from \texttt{secondary\_structure\_id()}; ranges indicate the first and last residue number of the secondary structure element (simplified DSSP: C, coil; H, helix). 
    In both cases $\phi$ and $\psi$ dihedral angles were used  as a projection.}
    \label{fig:section and ss}
\end{figure*}

On the other hand, \texttt{secondary\_structure\_id()}  function subdivides the protein sequence according to the DSSP algorithm, which assigns each residue to a secondary structure element based on geometric criteria. These sections are contiguous, non-overlapping, and variable in length, so the resulting ID values reflect both the intrinsic flexibility of each secondary structure element and a reduced influence from neighboring segments, as shown in Figure~\ref{fig:section and ss}(B), where secondary structure elements are defined on a reference folded structure (in this case the PDB structure used as input for the topology), ID shows  higher variability between sections than between folded and unfolded states, suggesting that local structural context rather than the folding state dominates the observed dimensionality. 

\subsection{Case Study: NTL9}
We repeated the analysis on another protein from the same dataset, the N-terminal domain of Ribosomal Protein L9 (NTL9, PDB: 2HBA(1-39)), a 39-residue protein carrying a single point mutation (K12M) that increases its stability by 1.9 kcal/mol \cite{ntl9}. This case was chosen to validate our package’s performance on a protein with a more complex topology (Supplementary Figure~\ref{fig:ntl9_topology}).  We selected the unbiased trajectory  \texttt{DESRES-Trajectory\_NTL9-2-protein}, evaluating its RMSD with respect to the folded structure (reference frame 12,500).  Following the  method described previously, we identified six segments of 200 ns each (2,000 frames), comprising three folded-state segments (\verb|f0|, \verb|f1|, \verb|f2|) and three unfolded-state segments (\verb|u0|, \verb|u1|, \verb|u2|) (Supplementary Table~\ref{tab: ntl9 trajectory}).

Consistent with villin results, the \textit{TwoNN} estimator distinguish between folded and unfolded states  (Supplementary Figure~\ref{ntl9 estimators comparison}). The  projection methods tested (carbon--carbon distances and torsional angles) have a similar trend, namely  the IDs computed with $\chi$ and Ramachandran angles shift in  opposite directions (Supplementary Figure \ref{fig:ntl9 projections}). In absolute terms, the ID metric yielded higher values for the folded trajectories.

\begin{figure}
    \centering
    \includegraphics[width=0.9\linewidth]{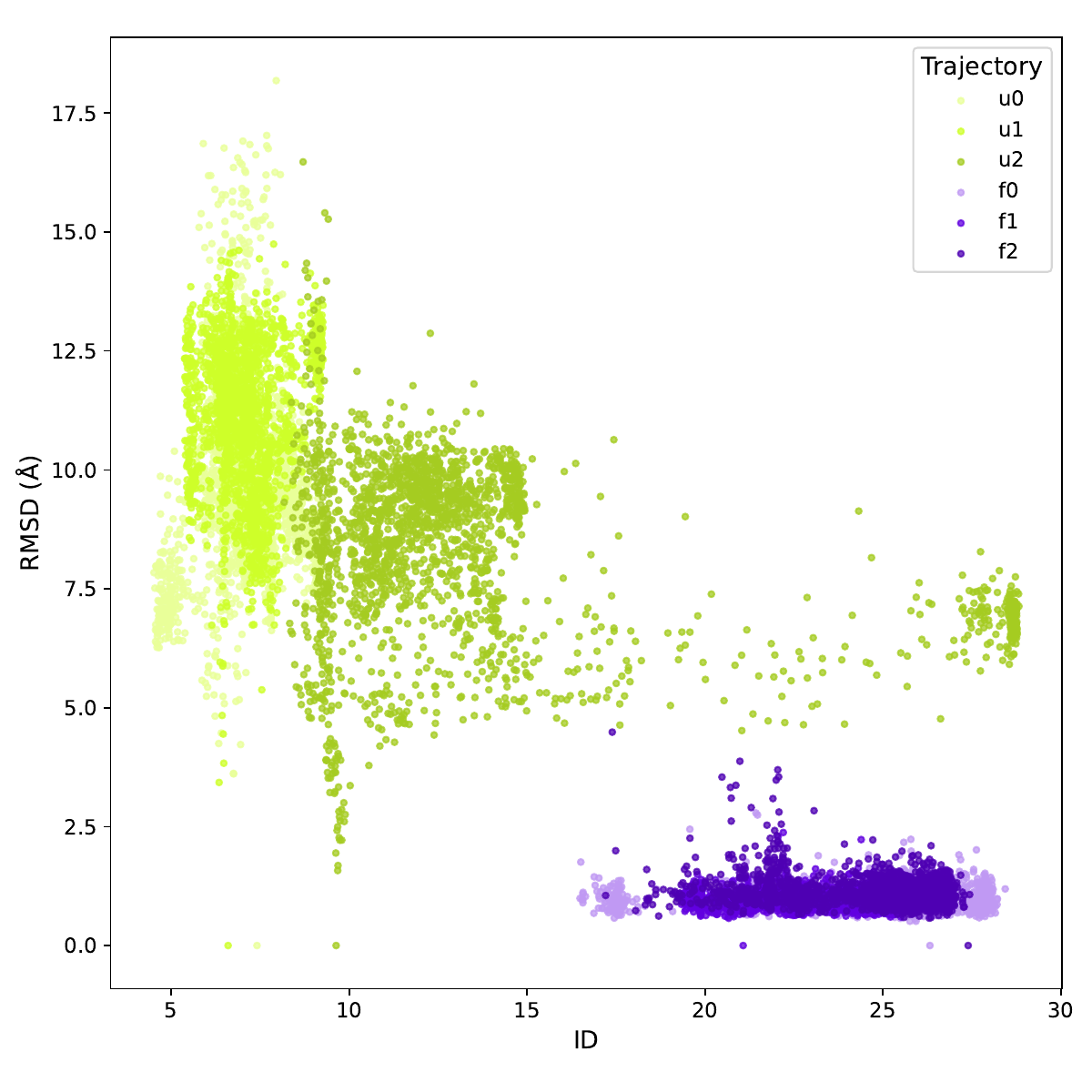}
    \caption{RMSD versus  instantaneous ID of NTL9 in folded (violet) and unfolded (green) states.
    Projection: $\phi$ and $\psi$ dihedral angles.}
    \label{fig:RMSD ntl9}
\end{figure}

RMSD-versus-ID analysis followed the expected anti-correlated trend (Figure~\ref{fig:RMSD ntl9}), with one noteworthy exception:  trajectory \texttt{u2} exhibited a period during which ID values were more typical of a folded segment, but with a high RMSD w.r.t.\ the folded structure. This combination indicates the detection of a transient non-native, but relatively stable, folding intermediate.
The intermediate, a three-helix globule (Supplementary Figure~\ref{fig:ntl9 u2}),  can be clearly identified with a peak in the \textit{instantaneous} ID between 160 ns and 180 ns (Supplementary Figure~\ref{fig:ntl9_instantaneous}), but not in the \textit{averaged} nor \textit{overall} ID metrics (Supplementary Figure~\ref{fig:ntl9 global id}).

Lastly, we evaluated NTL9 sequence and structure locality.  ID values derived from \texttt{secondary\_structure\_id()} (Supplementary Figure~\ref{fig:ntl9 section and ss}B), predominantly influenced by the structural context, were villin-like as expected. On the other hand, an analysis with  \texttt{section\_id()}    indicated that in trajectory \texttt{u2} N-terminal windows had higher ID values, indicating an increased folded character for region 1--30  (Supplementary Figure~\ref{fig:ntl9 section and ss}A).

\section{Discussion and Conclusions}

MD simulations produce high-dimensional datasets that are often difficult to interpret directly. Estimating the ID of conformational ensembles provides a compact measure of the effective degrees of freedom and offers a complementary view of molecular flexibility and structural heterogeneity. The package implements this idea using internal-coordinate representations and multiple estimators, enabling both trajectory-level and local, region-specific analyses.

\textcolor{\newtextcolor}{From a methodological standpoint, ID is closely related to, but distinct from, dimensionality estimates obtained from explicit embeddings such as PCA or tICA. Whereas these methods return an ordered set of collective coordinates and an associated explained-variance spectrum, ID summarizes in a single scalar the effective dimensionality of the manifold explored by the trajectory. A natural extension of the present work will be to systematically compare ID with the number of principal components or time-lagged independent components required to explain a given fraction (e.g., 90\%) of the variance, thereby helping practitioners connect this notion with more familiar linear analysis tools.}

When applied to unbiased folding–unfolding simulations, the method distinguishes folded, unfolded, and transiently folded states. These results show that intrinsic dimensionality reflects changes in dynamic heterogeneity and flexibility, thereby complementing established order parameters used in folding studies \cite{muff_identification_2009,wriggers_automated_2009}.

\textcolor{\newtextcolor}{We remark that an extended structure spends most of its time moving along a few soft collective directions (such as overall expansion and compaction), whereas a relatively compact globule supports many more, albeit smaller-amplitude, fluctuation modes. This reconciles the (apparently paradoxical) higher ID of the folded state with the intuitive picture that an unfolded chain has more freedom, as seen in  Section \ref{sec:results} with the case studies of villin and NTL9.}

More generally, estimating ID highlights collective dynamical regimes that may be obscured in the full coordinate space, and may inform the construction of data-driven collective variables and features for Markov state modeling \cite{nagel_selecting_2023,das_correlating_2024,tsai_sgoop_2021}.

The package's modular design, open-source availability, and compatibility with existing MD analysis workflows make it broadly applicable to diverse biomolecular systems, from proteins to nucleic acids and complexes. By bridging concepts from nonlinear data analysis and biophysical modeling, \texttt{MDIntrinsicDimension} hopefully offers a novel lens for exploring molecular flexibility and conformational landscapes.

\section*{Data and Software Availability}
The code is freely available from the GitHub repository  \href{https://github.com/giorginolab/MDIntrinsicDimension}{\url{giorginolab/MDIntrinsicDimension}}, together with extensive documentation and self-contained notebooks which reproduce figures and results of this paper. 
\textcolor{\newtextcolor}{The repository also includes regression tests and continuous-integration workflows that test the  analysis functions to ensure  stability as features and dependencies evolve.}  Data for  the case studies is available from ref.\ \cite{lindorff-larsen_how_2011}.

\section*{Competing Interests}

None declared.

\section*{Author Contributions Statement}

\textbf{IC}: Software, Methodology, Visualization. \textbf{TG}: Methodology, Conceptualization, Supervision. \textbf{All authors}: Writing -- Review \& Editing.

\section*{Acknowledgements}

The authors would like to thank D.\ E.\ Shaw Research for kindly providing the villin and NTL9 folding-unfolding trajectories. We acknowledge ISCRA for awarding this project access to the LEONARDO supercomputer, owned by the EuroHPC Joint Undertaking, hosted by CINECA (Italy).

\section*{Funding}

TG acknowledges funding from the Spoke 7 of the National Centre for HPC, Big Data and Quantum Computing (Centro Nazionale 01 CN0000013), funded by the European Union -- NextGenerationEU, Mission 4, Component 2, Investment line 1.4, CUP B93C22000620006; 
from the PRIN 2022 (BioCat4BioPol) from the Ministero dell'Università e Ricerca, funded by the European Union -- NextGenerationEU, Mission 4 Component C2, CUP B53D23015140006; 
from the Spoke 5 “Next-Gen Therapeutics” of PNRR M4C2 Investiment 1.3  “HEAL ITALIA” PE00000019, CUP H43C22000830006 project “PROPHECY-GlycoRARE”; and
from the project InvAt-Invecchiamento Attivo e in Salute (FOE 2022) CUP B53C22010140001. 

\section*{Supporting Information}

List of methods and their abbreviations, frame ranges used in the analysis, 
extended plots and analysis for villin HP35 and NTL9 trajectories.

\printbibliography

\end{document}


\title{Supporting Information \\[1cm] MDIntrinsicDimension: Dimensionality-Based Analysis of Collective Motions in Macromolecules from  Molecular Dynamics Trajectories }
\author{Irene Cazzaniga\textsuperscript{1}, Toni Giorgino\textsuperscript{1}}
\date{\small \textsuperscript{1}Istituto di Biofisica (IBF-CNR), Consiglio Nazionale delle Ricerche, Milano, Italy.}
\maketitle


\noindent This file contains the list of methods and their abbreviations, frame ranges used in the analysis, extended plots for villin HP35 and NTL9. The two-nearest neighbours (TwoNN) estimator was used for all plots unless indicated otherwise.


\section{Supporting Information for the Methods Section}

\begin{table*}[h!]
    \centering
    \begin{tabular}{p{9cm}ccc}
    \toprule
        \bf Estimator name & \bf Abbreviation & \bf Type & \bf Reference \\
    \midrule
        Correlation (fractal) dimensionality & CorrInt & local & \cite{grassberger_measuring_1983} \\
        Dimensionality from angle and norm concentration & DANCo  & global &  \cite{ceruti_danco_2014} \\ 
        Expected simplex skewness & ESS   & global & \cite{johnsson_low_2015} \\
        Fisher separability  & FisherS  & global & \cite{albergante_estimating_2019} \\
        Weighted average kNN distances & KNN  & local  & \cite{carter_local_2010} \\
        Local principal component analysis  & lPCA & local & \cite{fukunaga_algorithm_1971} \\
        Manifold-adaptive fractal dimension & MADA  & local & \cite{farahmand_manifold-adaptive_2007}\\
        Minimum neighbor distance--maximum likelihood & MiND\_ML  & global & \cite{rozza_novel_2012} \\
        Maximum likelihood estimator & MLE & local & \cite{levina_maximum_2005} \\
        Method of moments & MOM  & local & \cite{amsaleg_extreme-value-theoretic_2018} \\
        Tight localities estimator  & TLE  & global & \cite{amsaleg_intrinsic_2022} \\
        Two-nearest neighbours & TwoNN  & global & \cite{facco_estimating_2017} \\
    \bottomrule
    \end{tabular}
    \caption{ID estimators supported by the \texttt{scikit-dimension} library (alphabetically ordered).}
    \label{tab:scikit_estimators}
\end{table*}


\newpage

\section{Supporting Information for the villin HP35 Case Study}

\begin{table}[h!]
    \centering
    \begin{tabular}{cccc}
    \toprule
       \bf{TS}& \bf{Key} & \bf{Frames}& \bf{State} \\
    \midrule
       000  & u0  & 0--2000    & Unfolded  \\
       001  & f0  & 0--2000    & Folded   \\
       001  & u1  & 5000--7000 & Unfolded \\
       004  & f1  & 8000--10000& Folded   \\
       005  & u2  & 1700--3700 & Unfolded \\
       005  & f2  & 5000--7000 & Folded   \\
    \bottomrule
    \end{tabular}
    \caption{Subsections of the villin  trajectories selected for analyses, split into three  periods in the folded state and three  in the unfolded state. \textit{TS} is the trajectory segment, i.e. the dotted part in trajectory file names \texttt{2F4K-protein-0-\dots.dcd} from \cite{lindorff-larsen_how_2011}; \textit{Key} is the corresponding abbreviation used in the legends. }
    \label{tab: villin trajectory}
\end{table}

\begin{figure}[h!]
    \centering
    \includegraphics[width=0.7\linewidth]{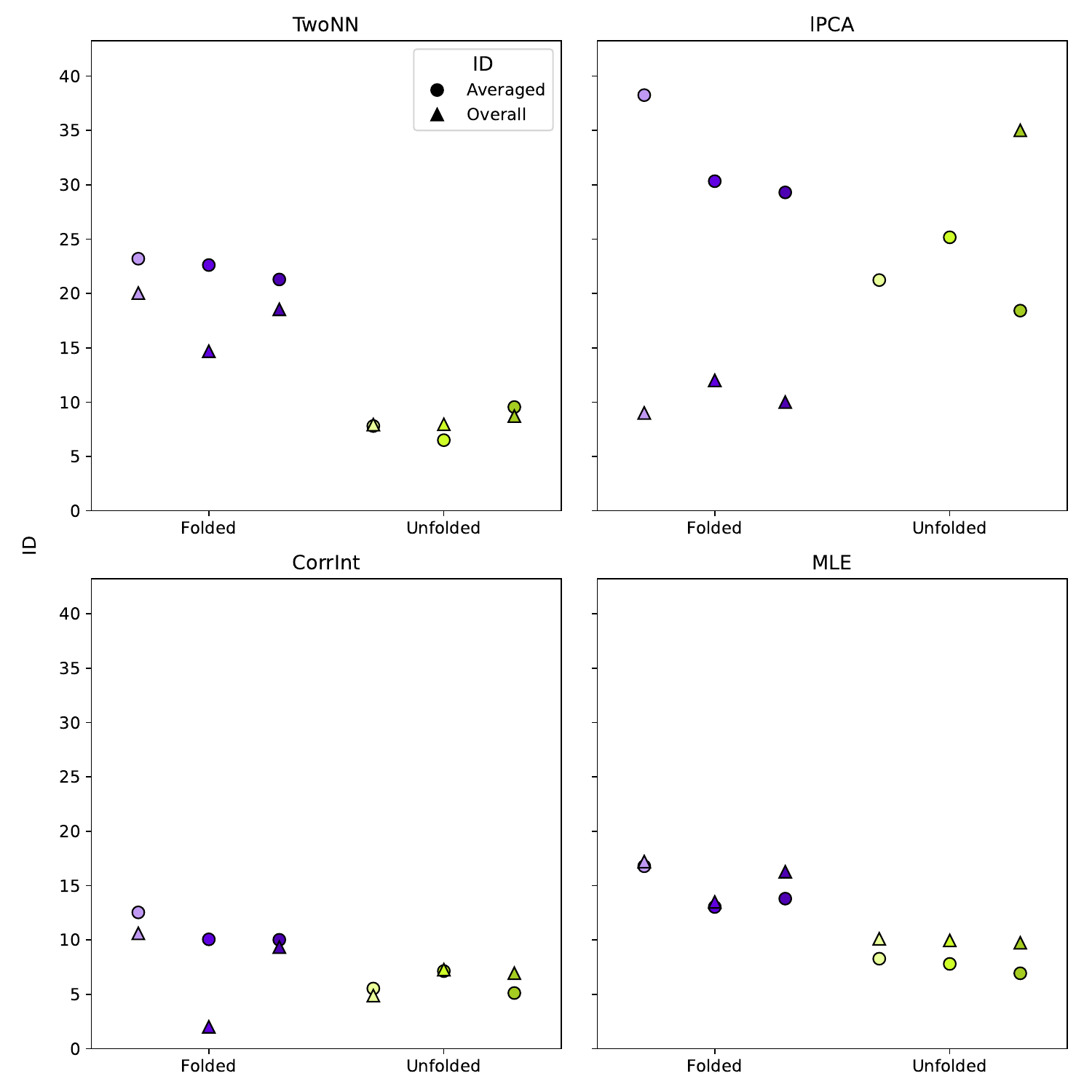}
    \caption{Averaged (circles) and overall (triangles) ID of the villin trajectories.
    The two summary metrics take on similar values for most estimators (only four are shown here for clarity).  lPCA is an exception but, as discussed in the main text, it does not discriminate states clearly. In all cases, $\phi$ and $\psi$ dihedral angles were used as a projection.}
    \label{fig:global id}
\end{figure}


\begin{figure}[h!]
    \centering
    \includegraphics[width = 1\linewidth]{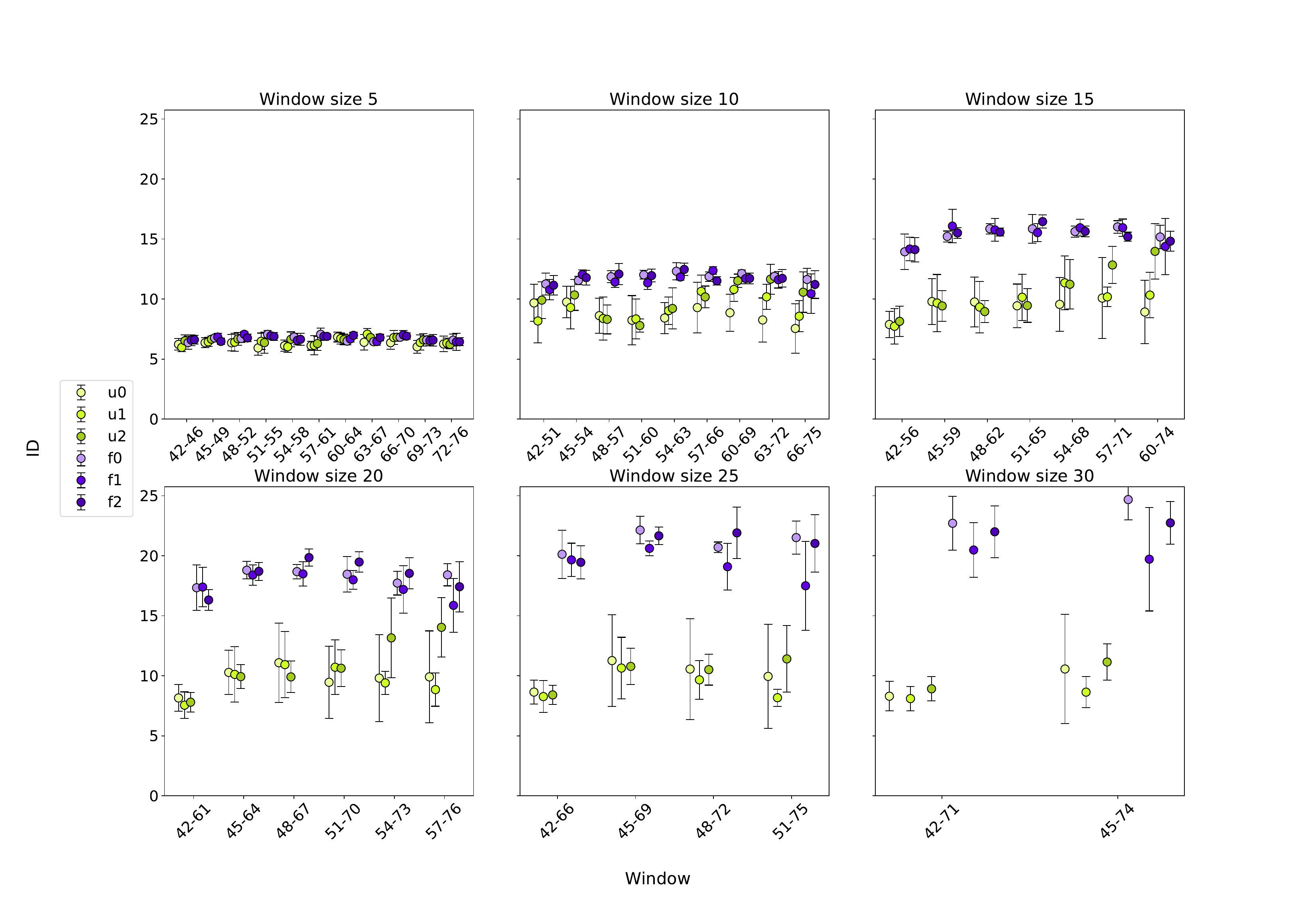}
    \caption{Sensitivity of section-wise ID (\texttt{section\_id()}) to the window size.  
    The window size strongly affects the ID, especially in case of folded states: for very small windows (e.g. 5) ID saturates and does not detect structural differences between the folded or unfolded states; at larger window sizes the classification becomes clearly distinct. Even larger window sizes blur sequence details to the point where the ID approaches the whole-protein value (e.g. window size of 30, with 35 total residues).}
    \label{fig:villin windows}
\end{figure}

\begin{figure}[h!]
    \centering
    \includegraphics[width = 0.9\linewidth]{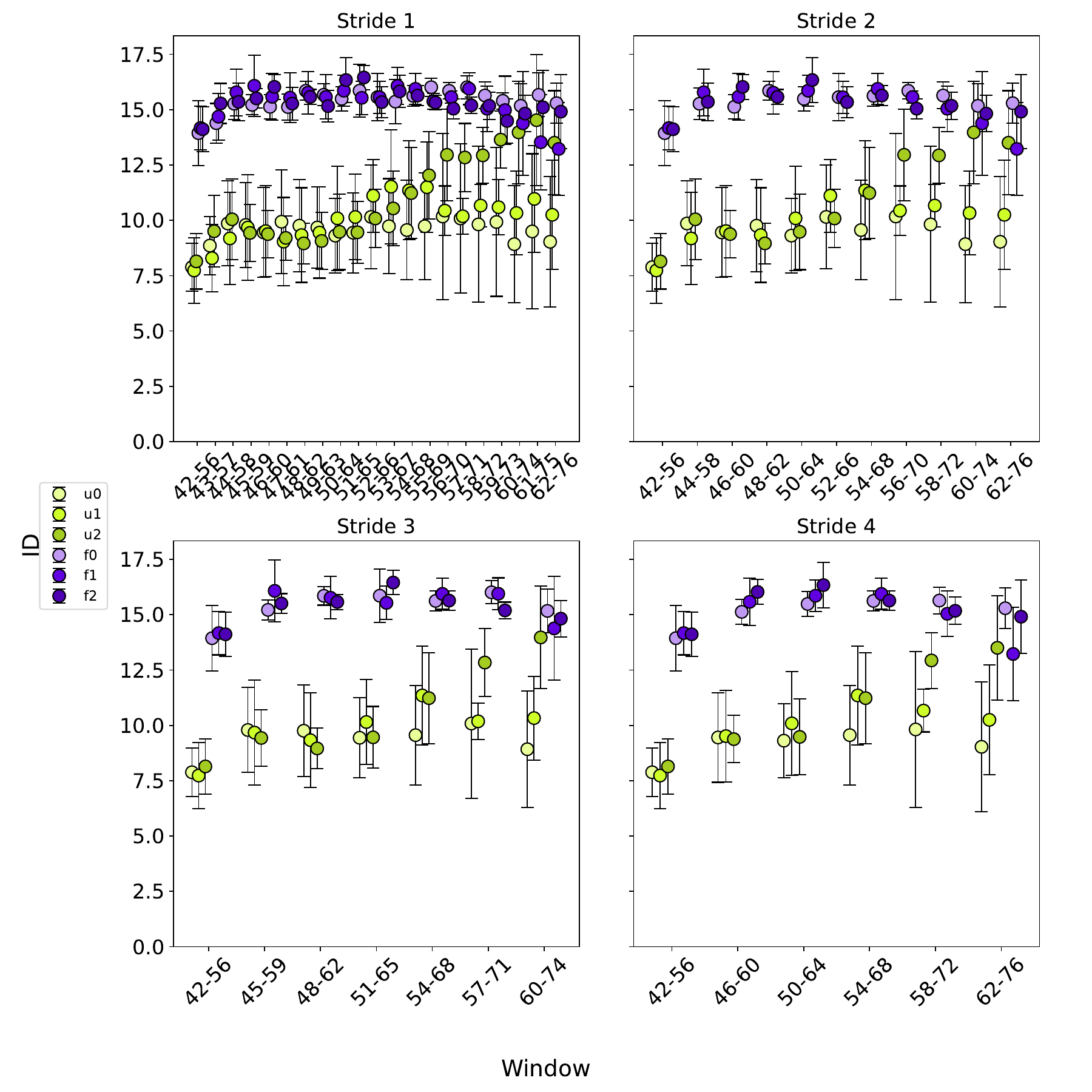}
    \caption{Sensitivity of section-wise ID (\texttt{section\_id()}) to the stride. Expectedly, a large stride makes the ID profiles sparser without altering their trend.}
    \label{fig:villin strides}
\end{figure}

\begin{figure}[h!]
    \centering
    \includegraphics[width=0.49\linewidth]{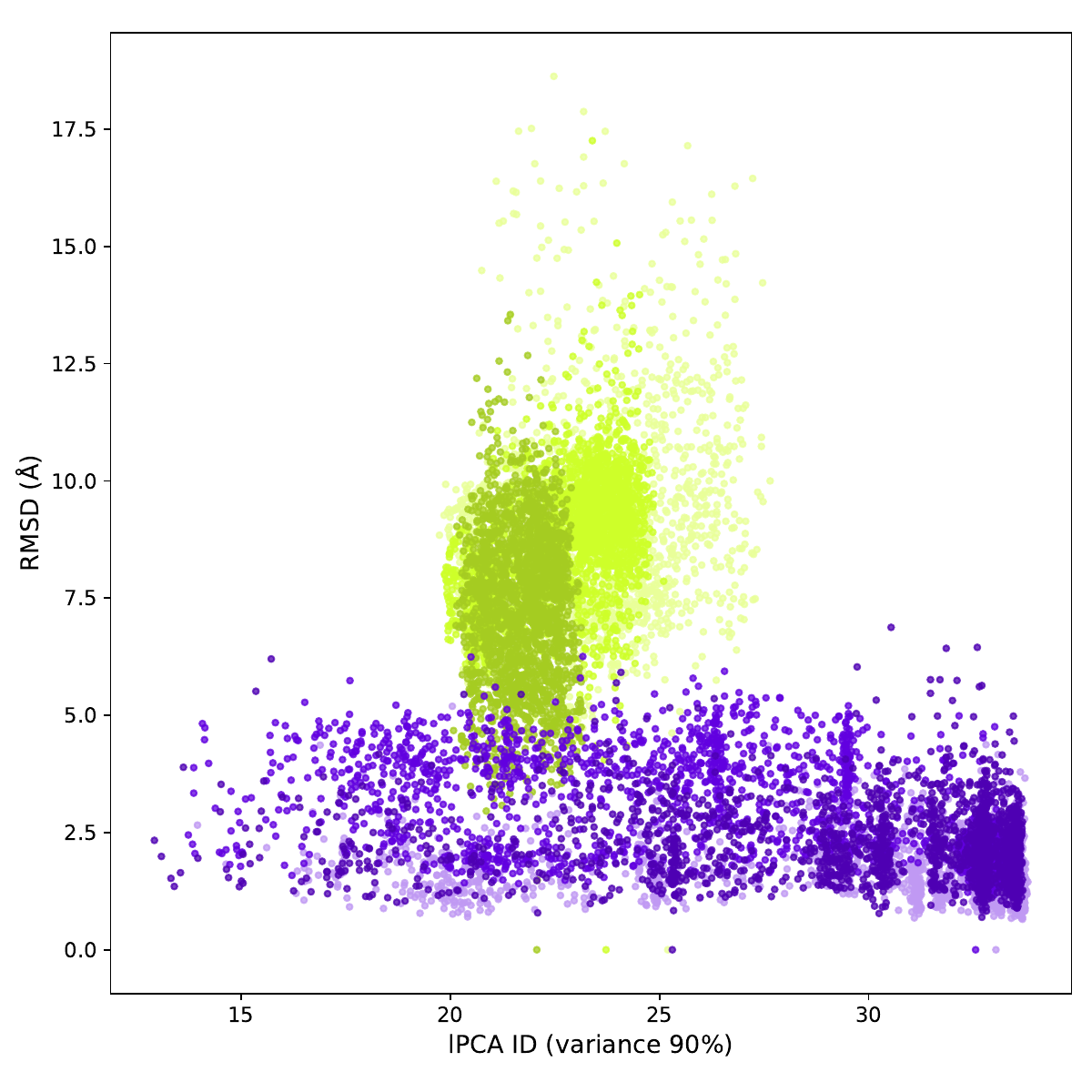}
    \includegraphics[width=0.49\linewidth]{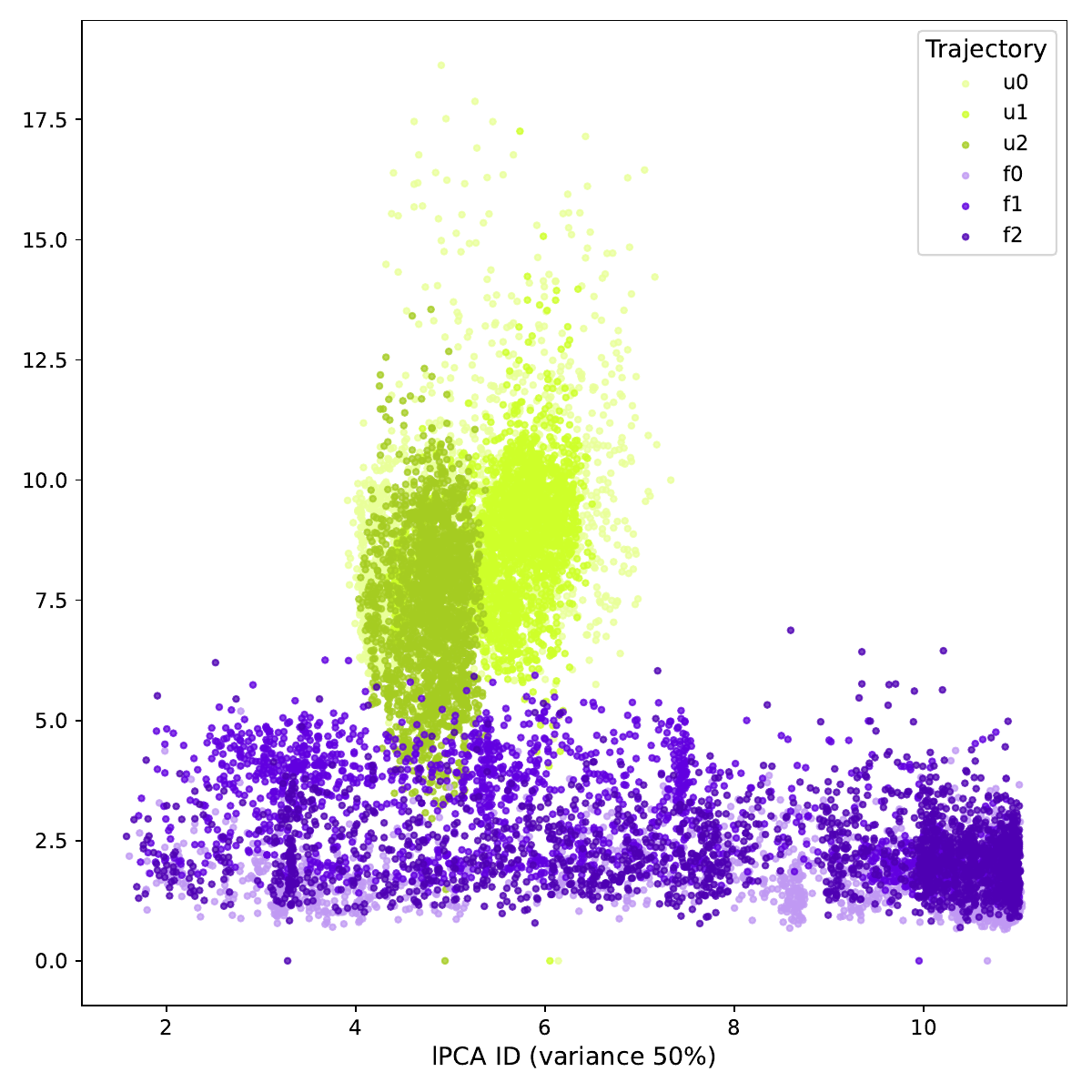}
    \caption{Comparison between PCA-based ID estimation and RMSD on MD simulations of villin. The horizontal axis corresponds to the instantaneous ID estimated as the number of eigenvalues accounting for 90\%\ (left) or 50\%\ (right) of the local variance. Both  thresholds fail to separate the folded (violet) and unfolded (green) states trajectories. Hence, ID computed via modern, nonlinear estimators like TwoNN provide a more precise estimate. 
    Projection: $\phi$ and $\psi$ dihedral angles.}
    \label{fig:villin lpca rmsd}
\end{figure}

\begin{figure}[h!]
    \centering
    \includegraphics[width=0.9\linewidth]{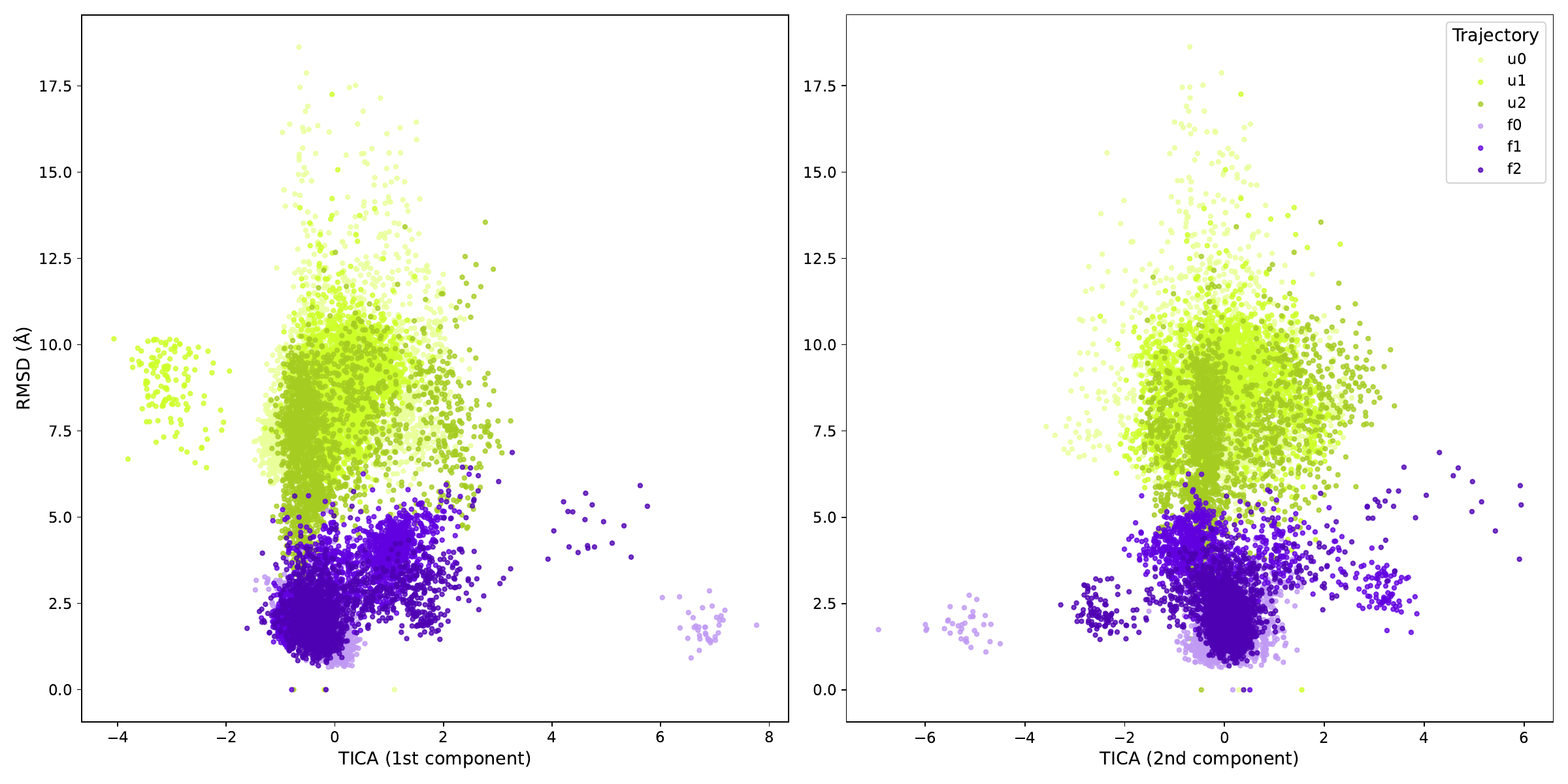}
    \caption{Comparison between tICA (first and second components) and RMSD on MD simulations of villin. Neither of the two tICA components discriminates the folded (violet) vs. unfolded (green) trajectories.
    Projection: $\phi$ and $\psi$ dihedral angles.}
    \label{fig:villin tica rmsd}
\end{figure}


\newpage
\clearpage

\section{Supporting Information for the NTL9 Case Study}

In the following, ``NTL9'' refers to the N-terminal
Domain of Ribosomal Protein L9 (NTL9, PDB:2HBA(1-39)). Trajectories refer to \texttt{DESRES-Trajectory NTL9-2-protein} dataset from the D.\ E.\ Shaw Research fast-folding proteins \cite{lindorff-larsen_how_2011}.
If not stated otherwise, the estimator of choice is \textit{TwoNN} \cite{facco_estimating_2017}, and the projections are computed from $\phi$ and $\psi$ Ramachandran angles.

\begin{figure}[h!]
\centering
\includegraphics[width=0.9\linewidth]{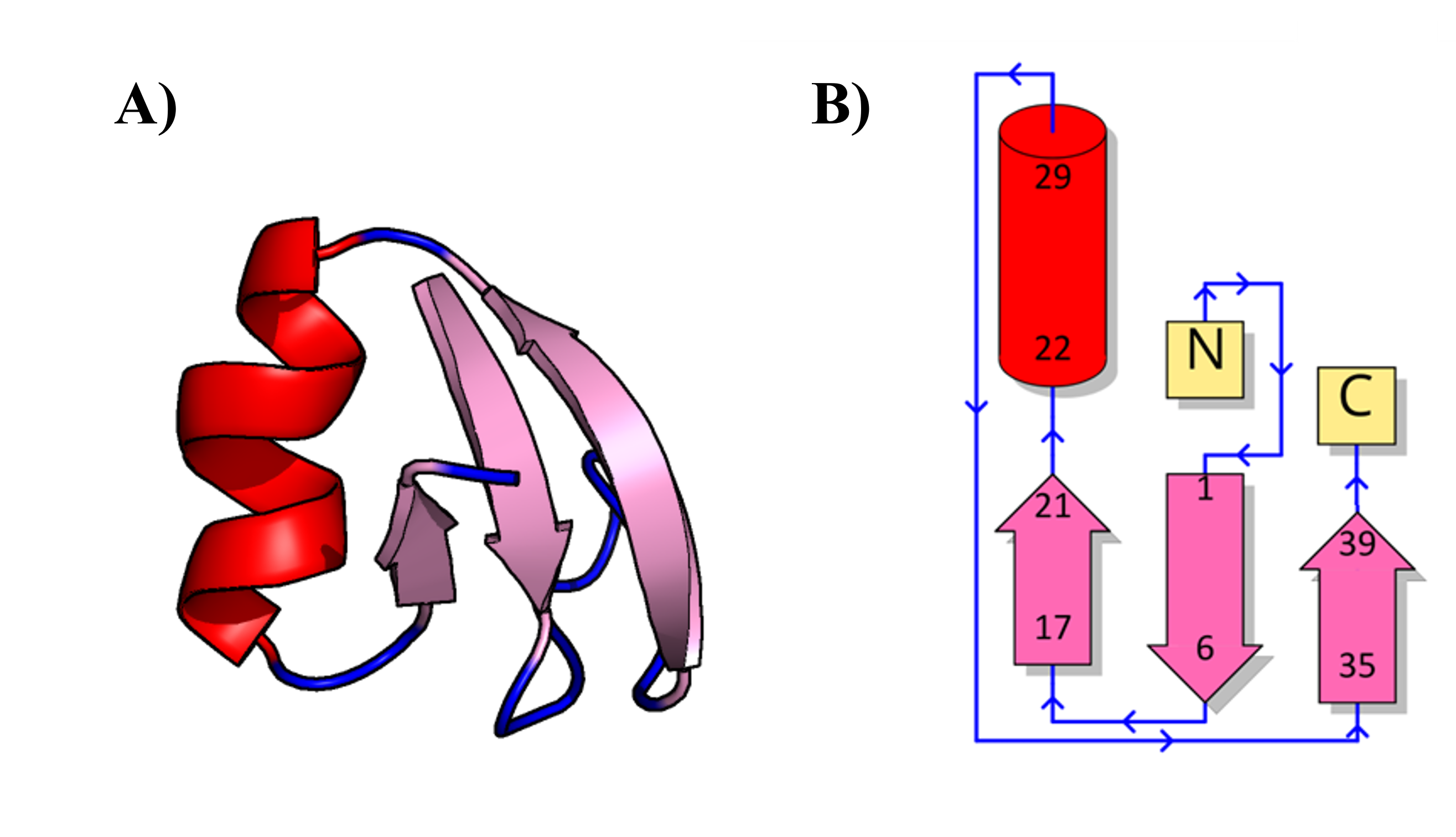}
    \caption{NTL9 (PDB 2HBA), 39-residue protein structure (A), colored by secondary structure and corresponding topology diagram (B).}
    \label{fig:ntl9_topology}
\end{figure}

\begin{table*}[h!]
    \centering
    \begin{tabular}{llll}
    \toprule
    \bf{Traj}& \bf{Key}& \bf{Frames}  & \bf{State}\\
    \midrule
    000 & u0 & 0--2000 & Unfolded \\
    050 & f0 & 0--2000 & Folded   \\
    080 & u1 & 0--2000 & Unfolded \\
    100 & f1 & 0--2000 & Folded   \\
    124 & u2 & 0--2000 & Unfolded \\
    194 & f2 & 0--2000 & Folded   \\
    \bottomrule
    \end{tabular}
    \caption{Subsections of the NTL9 trajectories selected for analyses, sliced into three periods in the folded state and three in the unfolded state. ``Traj'' correspond the trailing part of data files names \texttt{NTL9-2-protein-\dots.dcd} and accompanying paper~\cite{lindorff-larsen_how_2011}; ``Key'' is the corresponding abbreviation used in the legends.}
    \label{tab: ntl9 trajectory}
\end{table*}

\begin{figure}[h!]
    \centering
    \includegraphics[width=0.9\linewidth]{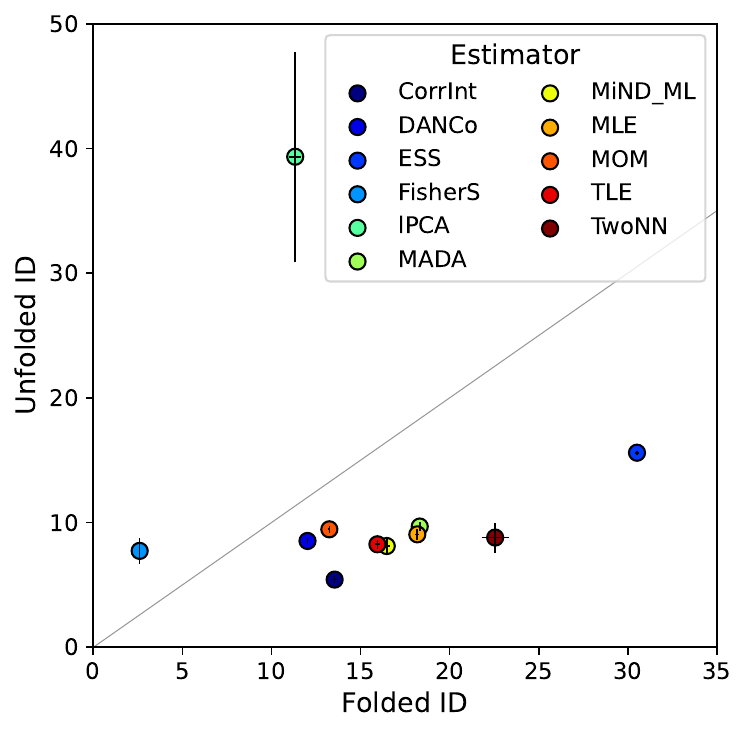}
    \caption{Folded--vs--unfolded ID of the NTL9 dynamic manifold, computed bu the estimators available at \texttt{scikit-dimension} package. Each point represents the mean value of the folded states ($x$ axis) and unfolded states ($y$ axis). Error bars parallel to each axis indicate the standard deviation from the mean for the respective state.}
    \label{ntl9 estimators comparison}
\end{figure}

\begin{figure}[h!]
    \centering
    \includegraphics[width=0.9\linewidth]{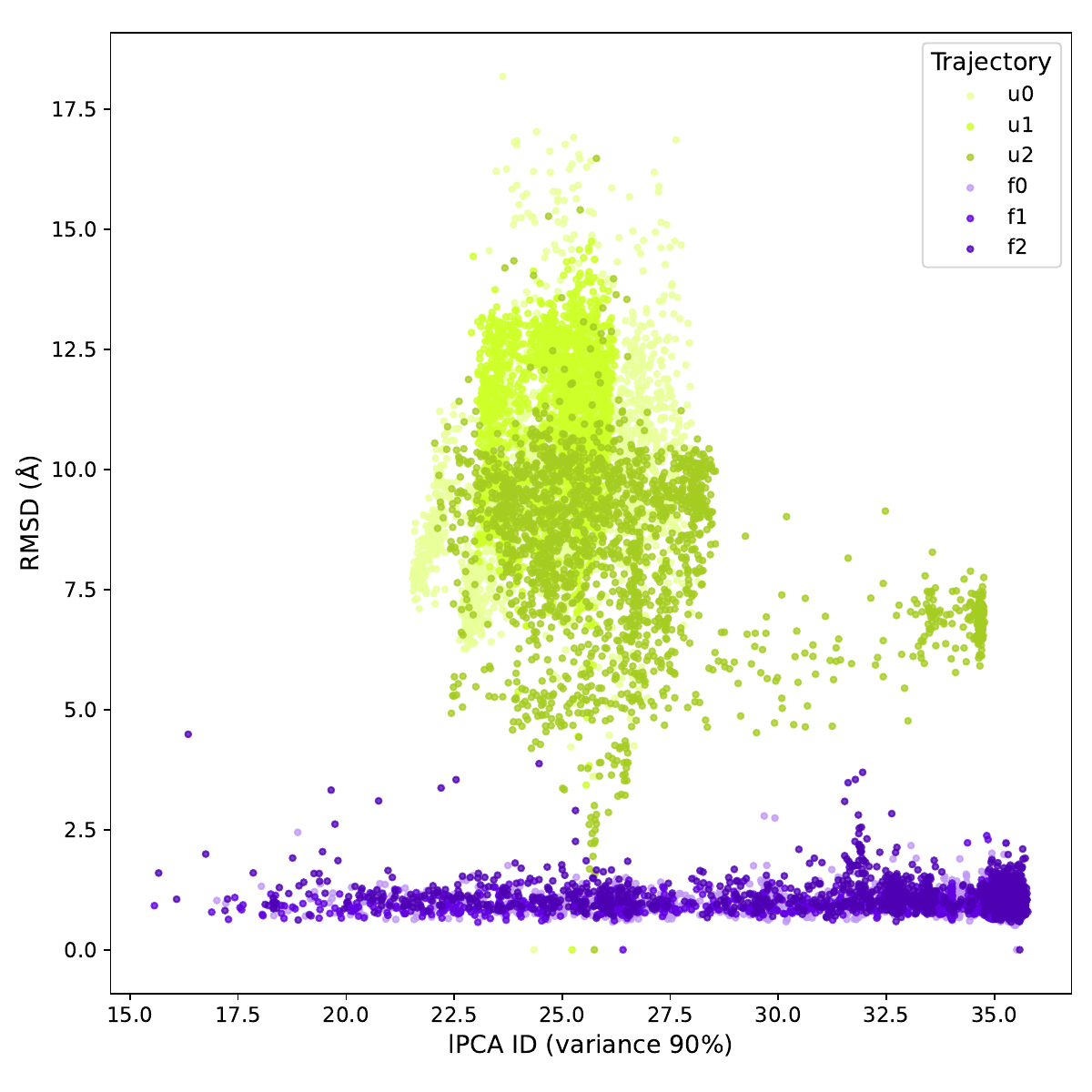}
    \caption{RMSD versus lPCA of NTL9 in folded (violet) and unfolded (green) states. Projection: $\phi$ and $\psi$ dihedral angles. Refer to caption of Supplementary Figure \ref{fig:villin lpca rmsd} (left) for details. Similarly to villin, lPCA does not discriminate the folding state.}
    \label{fig:ntl9 lpca rmsd}
\end{figure}

\begin{figure}[h!]
    \centering
    \includegraphics[width=0.9\linewidth]{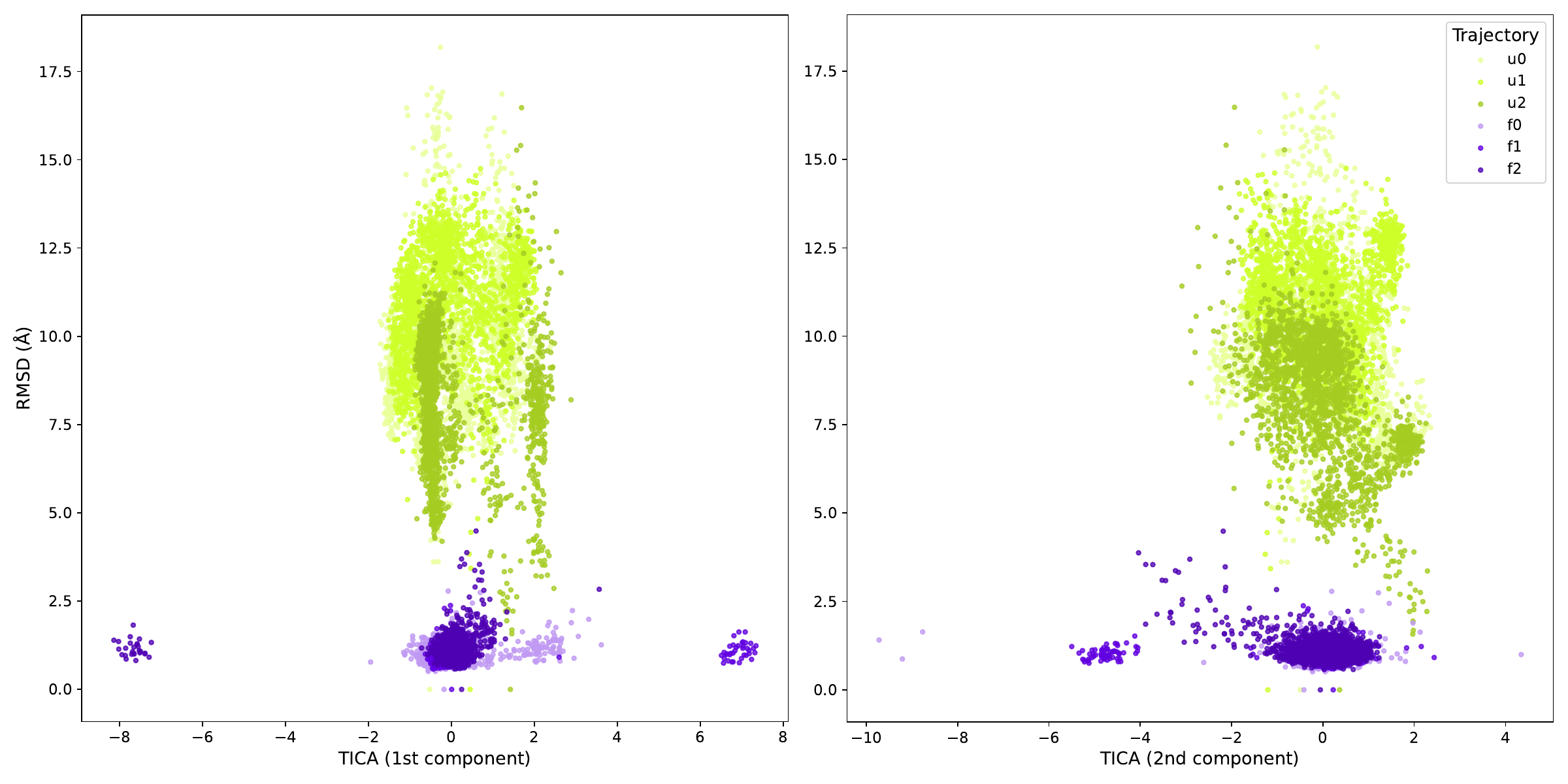}
    \caption{RMSD versus tICA components 1 and 2 of NTL9 in folded (violet) and unfolded (green) states. Projection: $\phi$ and $\psi$ dihedral angles.
    Refer to caption of Supplementary Figure \ref{fig:villin tica rmsd} for details.}
    \label{fig:ntl9 tica rmsd}
\end{figure}

\begin{figure}[h!]
    \centering
    \includegraphics[width=0.9\linewidth]{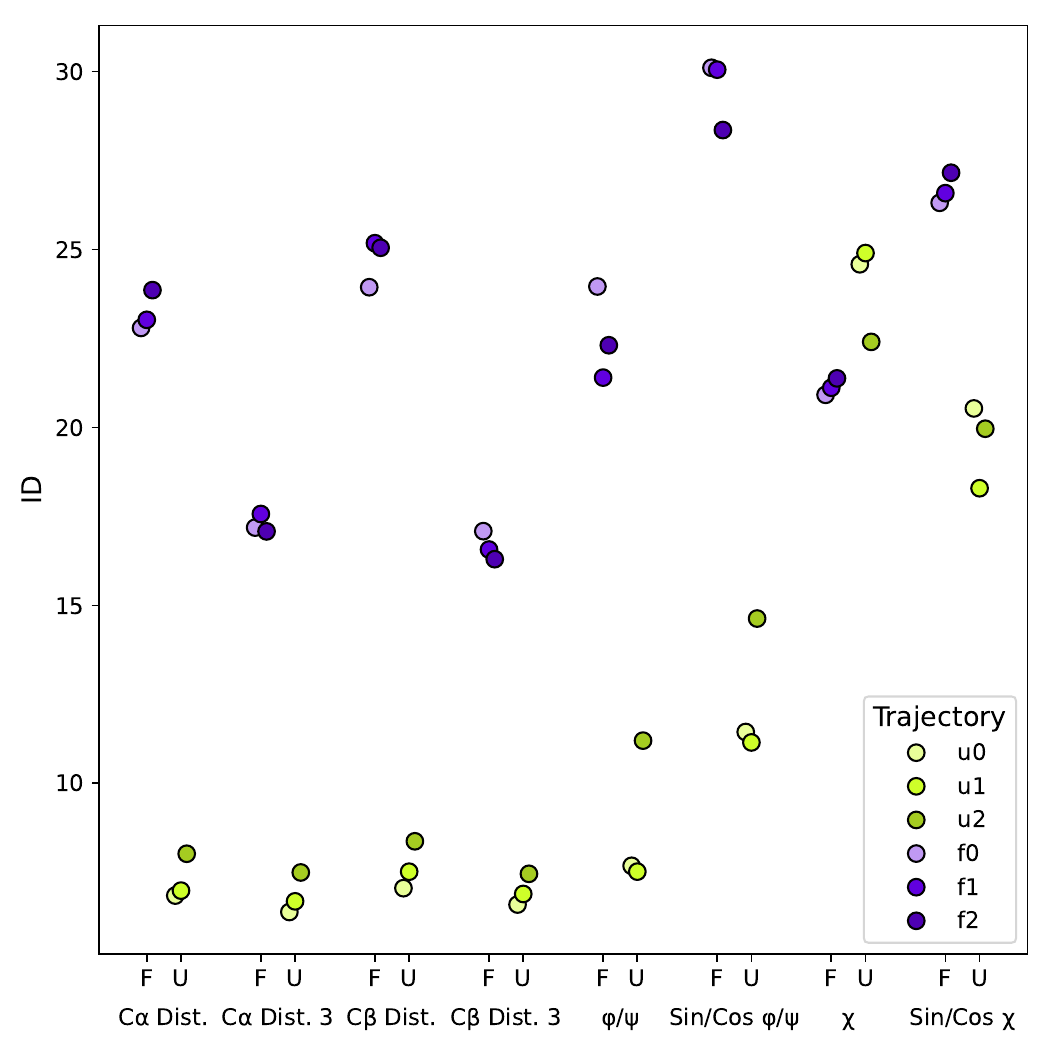}
    \caption{Differences in ID between folded (f) and unfolded (u) states on NTL9 under different projections. \textit{Dist.}: pairwise distances between all carbon--carbon pairs; \textit{Dist. 3}: pairwise distances every 3rd carbon; $\phi, \psi$: Ramachandran angles; $\chi$: sidechain dihedrals; \textit{Sin/Cos}: trigonometric embedding of angles.}
    \label{fig:ntl9 projections}
\end{figure}

\begin{figure}[h!]
    \centering
    \includegraphics[width=0.5\linewidth]{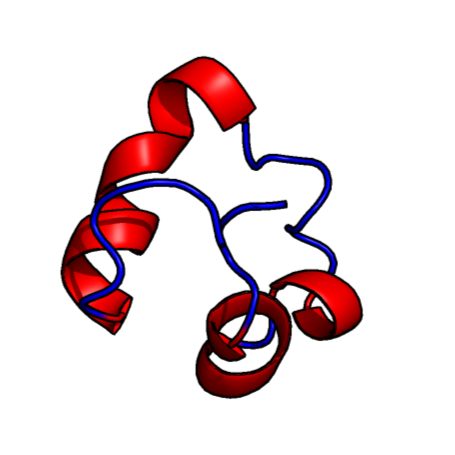}
    \caption{Transiently folding intermediate of NTL9 found in trajectory \texttt{u2} at frame 1700. }
    \label{fig:ntl9 u2}
\end{figure}

\begin{figure*}[h!]
    \centering
    \includegraphics[width=0.9\linewidth]{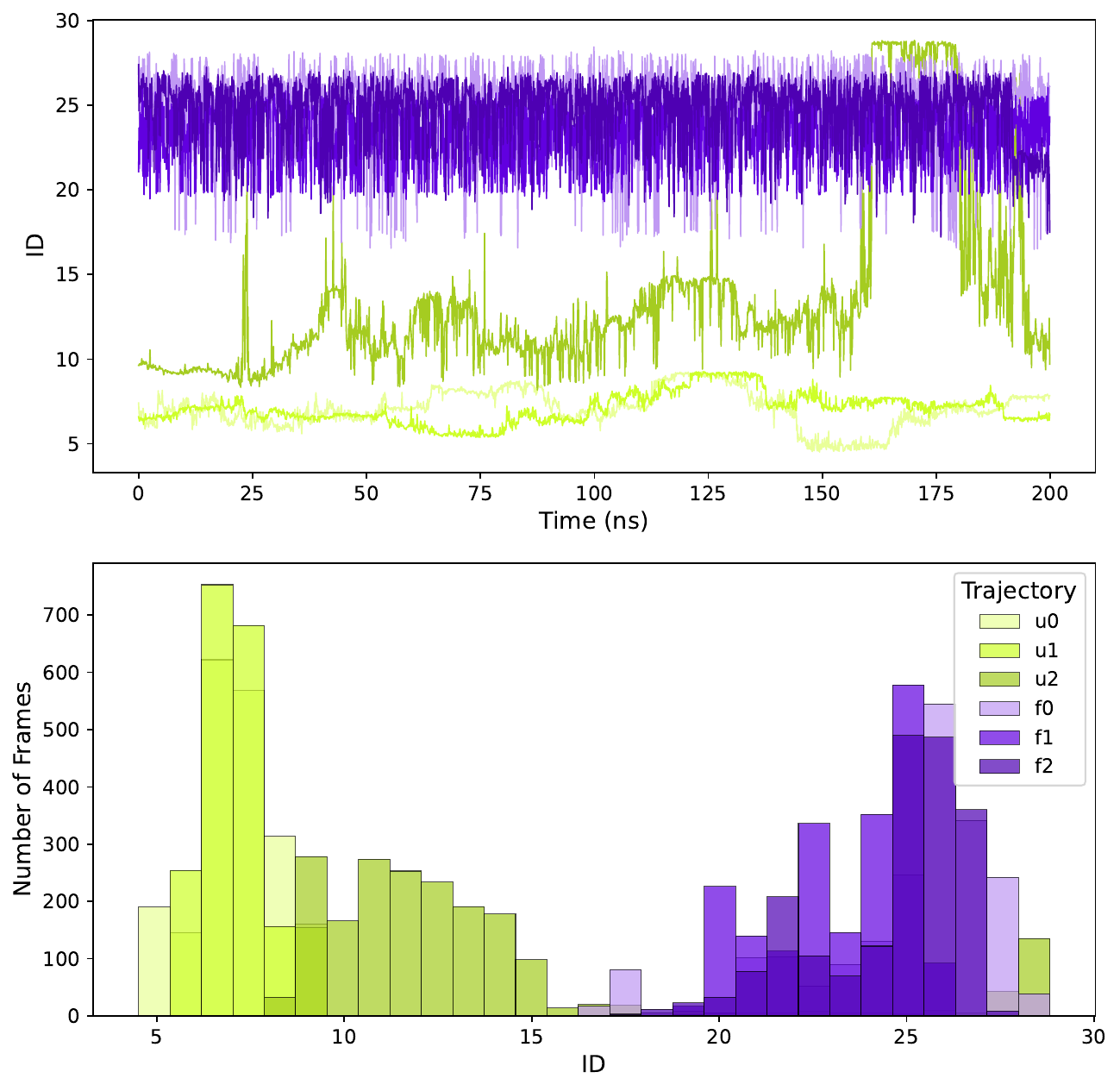}
    \caption{Instantaneous ID over time (top panel) and ID distribution frequency (bottom panel) along the trajectory.
    In violet, folded protein trajectories; in green, unfolded ones.}
    \label{fig:ntl9_instantaneous}
\end{figure*}

\begin{figure}[h!]
    \centering
    \includegraphics[width=0.6\linewidth]{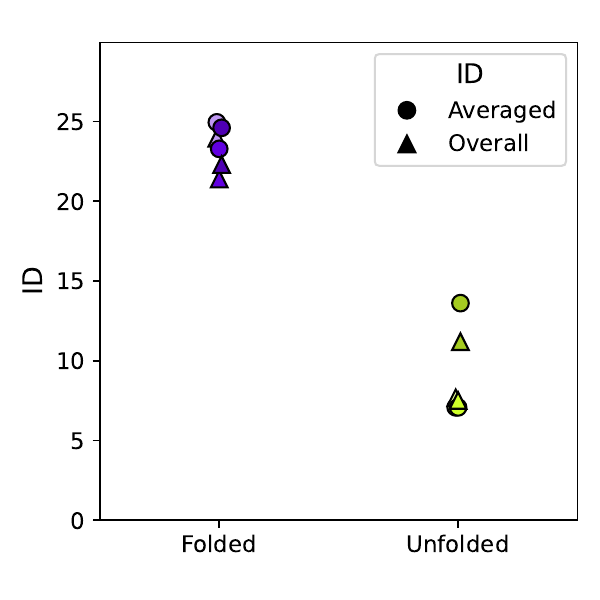}
    \caption{Averaged (circles) and overall (triangles) ID from individual trajectories. }
    \label{fig:ntl9 global id}
\end{figure}

\begin{figure*}[h!]
    \centering
    \includegraphics[width=\linewidth]{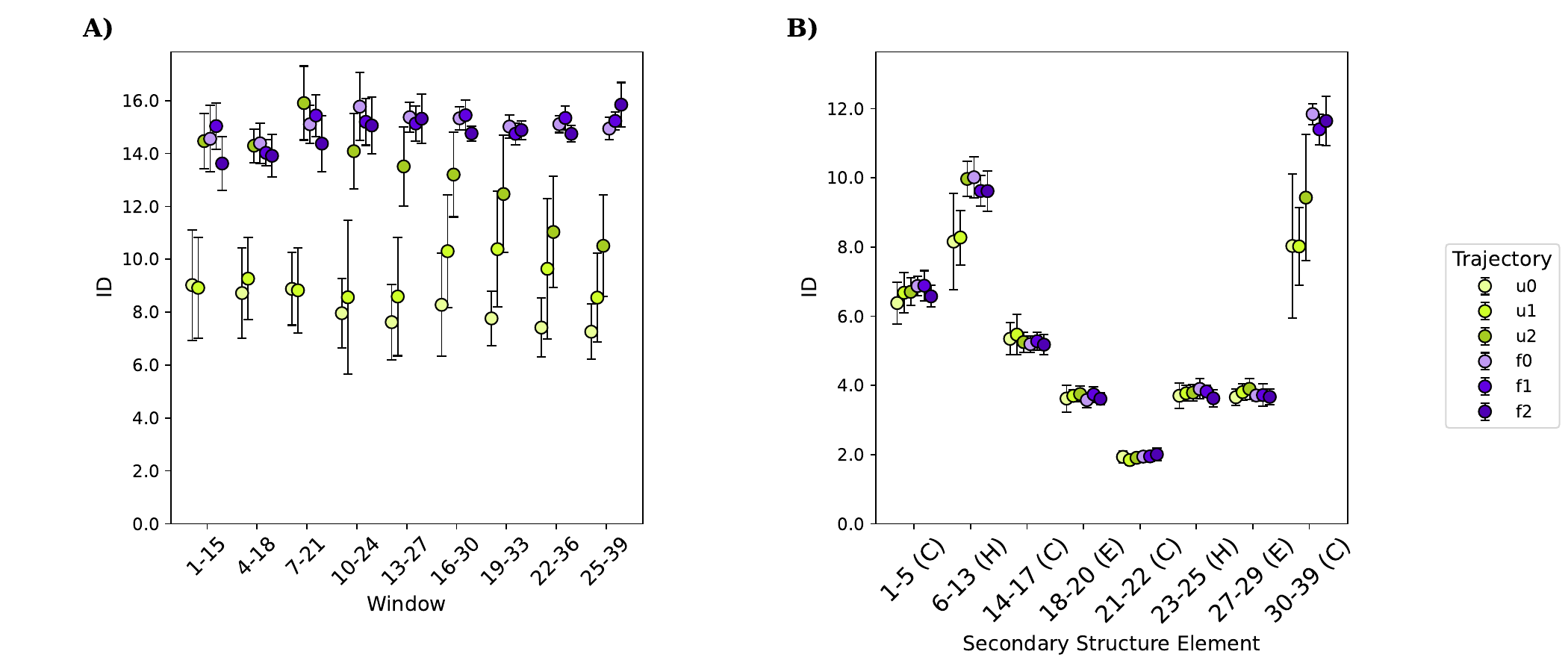}
    \caption{ID computed locally and on structural  regions: (A) Sequence-wise ID computed with \texttt{section\_id()} using a \textit{window} of 15 and \textit{stride} of 3; (B) Structure-wise ID computed with \texttt{secondary\_structure\_id()}, based on  simplified 3-class DSSP assignments.}
    \label{fig:ntl9 section and ss}
\end{figure*}

\clearpage

\bibliographystyle{vancouver}
\bibliography{bibliography}